\documentclass[11pt,onecolumn,amssymb,nofootinbib]{revtex4}
\usepackage{amsmath, amsthm, amscd, amssymb}
\usepackage{bm}
\usepackage{bbm}

\begin{document}

\title{\bf Covariance Group for Null Geodesic Expansion Calculations, and its Application to the Apparent Horizon} \bigskip
\leftline{Essay written for the Gravity Research Foundation 2021 Awards for Essays on Gravitation;}
\leftline{submitted March 30, 2021.}

\author{Stephen L. Adler}
\email{adler@ias.edu} \affiliation{Institute for Advanced Study,
1 Einstein Drive, Princeton, NJ 08540, USA.}

\begin{abstract}
We show that the recipe for computing the expansions $\theta_\ell$ and $\theta_n$ of outgoing and ingoing null geodesics normal to a surface   admits a covariance group with  nonconstant scalar $\kappa(x)$, corresponding to the mapping $\theta_\ell \to \kappa \theta_\ell$, $\theta_n \to \kappa^{-1} \theta_n$. Under this mapping, the product $\theta_\ell \theta_n$ is invariant, and thus the marginal surface computed from the vanishing of $\theta_\ell$, which is  used to define the apparent horizon,  is invariant.   This covariance group naturally appears in comparing the expansions  computed with different choices of coordinate system.
\end{abstract}

\maketitle

In speaking of the horizon of a black hole, one usually thinks of the {\it event horizon}, which is a surface dividing spacetime into two causally disconnected regions.  Because identifying the event horizon requires knowing the entire causal structure of spacetime, it is feasible only for certain exactly computable black hole geometries, with  the static Schwarzschild and Kerr metrics being prime examples.  However, in rapidly evolving situations, such as numerical calculations  of black hole collisions leading to production of gravitational waves, the entire causal structure of spacetime is not attainable.  So instead one employs the concept of the {\it apparent horizon}, defined as the outermost marginally trapped surface \cite{hawking}.

The standard recipe for locating apparent horizons proceeds as discussed in the reviews \cite{faraoni}, \cite{krishnan1}.  For a  compact, orientable  2-surface embedded in 4-space, there are two orthogonal directions corresponding to outgoing and ingoing null rays, with respective tangents $\ell^{\nu}$ and $n^{\nu}$ respectively.  Let $g_{\mu\nu}$ be the metric \big(for which we take the (-,+,+,+) convention\big) and $g^{\mu\nu}$ its inverse. We begin by forming the projector onto the 2-surface,
\begin{equation}\label{proj1}
h^{\mu\nu}=g^{\mu\nu}+\frac{\ell^{\mu}n^{\nu}+\ell^{\nu}n^{\mu}}{-\ell^{\alpha}n_{\alpha}}~~~,
\end{equation}
which if we adopt \cite{nielsen} the convenient normalization convention
\begin{equation}\label{normconv}
\ell^{\alpha}n_{\alpha}=-2~~~,
\end{equation}
becomes
\begin{equation}\label{proj2}
h^{\mu\nu}=g^{\mu\nu}+\frac{\ell^{\mu}n^{\nu}+\ell^{\nu}n^{\mu}}{2}~~~.
\end{equation}
By construction,  $h^{\mu\nu}$ projects the null vector normals to zero,
\begin{equation}\label{zero}
h^{\mu\nu}\ell_\mu =h^{\mu\nu}\ell_\nu =h^{\mu\nu}n_\mu =h^{\mu\nu}n_\nu = 0~~~.
\end{equation}
Evidently in either formulation, the projector $h^{\mu\nu}$ is invariant under reciprocal rescalings of $\ell^{\nu}$ and $n^{\nu}$ according to
\begin{equation}\label{rescaling}
\ell^\nu \to \kappa \ell^\nu~,~~~n^\nu \to \kappa^{-1} n^\nu~~~.
\end{equation}
A rescaling of $\ell^\nu$ and $n^{\nu}$ with constant $\kappa$ has been considered previously by Ashtekar, Beetle, and Lewandowski \cite{ashtekar},  Ashtekar and Krishnan \cite{krishnan2} , and Krishnan \cite{krishnan1}, but here we allow $\kappa=\kappa(x)$ to be a general
nonconstant scalar function of the spacetime coordinate $x$.

The {\it expansion}  $\theta_{\ell}$ of a bundle (or congruence) of null rays associated with the tangent vector $\ell$  is a measure of the fractional change of the cross sectional area of the bundle as one moves along the central ray of the bundle.  Using the projector $h^{\mu\nu}$, the expansions $\theta_\ell$ and $\theta_n$ associated with the outgoing and ingoing null vectors $\ell^{\nu}$ and $n^{\nu}$ are calculated from the formula
\begin{align}\label{expansion}
\theta_\ell=&h^{\mu\nu} \nabla_\mu\ell_\nu~~~,\cr
\theta_n=&h^{\mu\nu} \nabla_\mu  n_\nu~~~.\cr
\end{align}
To see how these expansions transform under the rescalings of Eq. \eqref{rescaling}, we note that
\begin{equation}\label{nabla1}
\nabla_\mu \kappa(x) \ell_\nu = \ell_\nu \partial_\mu \kappa(x) + \kappa(x) \nabla_\mu \ell_\nu~~~.
\end{equation}
Since the inhomogeneous term $\ell_\nu \partial_\mu \kappa $ is projected to zero by $h^{\mu\nu}$ by virtue of Eq. \eqref{zero}, the expansion
$\theta_{\ell}$ transforms under Eq. \eqref{rescaling} by the simple scaling formula
\begin{equation}\label{thetatrans1}
\theta_\ell \to \kappa \theta_\ell~~~,
\end{equation}
and similarly, $\theta_n$ transforms under the reciprocal scaling formula
\begin{equation}\label{thetatrans2}
\theta_n \to \kappa^{-1} \theta_n~~~,
\end{equation}
with the product $\theta_\ell \theta_n$ invariant
\begin{equation}\label{prodtrans}
\theta_\ell \theta_n \to \theta_\ell \theta_n~~~.
\end{equation}
Thus calculations of the expansions $\theta_\ell$, $\theta_n$ using the standard recipe have a covariance group under the transformations of Eq. \eqref{rescaling} with general non-constant $\kappa(x)$, with the associated product of Eq. \eqref{prodtrans} an invariant.

Consider now what happens if we compute the expansions $\theta_{\ell,n}$ for the same physics viewed from different choices of coordinates.
In each coordinate system we have to pick null vectors $\ell$ and $n$, and different ways of doing this that satisfy the norm convention of
Eq. \eqref{normconv} will differ by the rescaling freedom of Eq. \eqref{rescaling}.  Thus, if we pick the most convenient definitions of $\ell$ and $n$ in each coordinate system, for example those with equal time components and opposite signs of the spatial components, we will in general get {\it different} values of the expansions $\theta_\ell$, $\theta_n$ in the various coordinate systems, with only the product $\theta_\ell \theta_n$ the same in all systems.  However, the value of this product is in itself a useful diagnostic.  According to the usual classification reviewed in \cite{faraoni}--\cite{nielsen}, $\theta_\ell \theta_n<0$ corresponds to a normal or untrapped surface; $\theta_\ell \theta_n>0$ corresponds to a trapped or antitrapped surface; and $\theta_\ell\theta_n=0$ corresponds to a marginal surface, such as the case $\theta_\ell=0$, $\theta_n>0$ which defines the future apparent horizon.  Thus an apparent horizon can be located by computing $\theta_\ell \theta_n$ in any coordinate system, even though the individual values of $\theta_\ell$ and $\theta_n$ may vary.

We illustrate this by the simple example of a static, spherically symmetric spacetime, viewed either from Gullstrand--Painlev\'e (GP) or from spherical (S)  coordinates.  The general static, spherically symmetric line element in GP coordinates is
\begin{equation}\label{GPelt}
ds^2=-[c(r)^2-v(r)^2]dt^2+2v(r)dr dt  + dr^2 +r^2d\Omega^2~~~,
\end{equation}
corresponding to $g_{tt}=-(c(r)^2-v(r)^2)$, $g_{rt}=g_{tr}=v(r)$, and $g_{rr}=1$.
Making the convenient and symmetrical choice \cite{nielsen}
\begin{align}\label{nv1}
\ell^\nu=&\big(1,c(r)-v(r),0,0\big)/c(r) ~~~,\cr
n^\nu=&\big(1,-c(r)-v(r),0,0\big)/c(r)  ~~~,\cr
\end{align}
which as required obey $\ell^\alpha  \ell_\alpha = n^\alpha  n_\alpha =0$ and $\ell^\alpha   n_\alpha =-2$,  a calculation \cite{nielsen} using the recipe of Eq. \eqref{expansion} (which can be checked by
 Mathematica) gives for the expansions
 \begin{align}\label{GPexpan}
 \theta_{\ell}=& 2\frac{\big(c(r)-v(r)\big)}{rc(r)}~~~,\cr
 \theta_{n}=& -2\frac{\big(c(r)+v(r)\big)}{rc(r)}~~~,\cr
 \theta_\ell \theta_n =& -4\frac{\big(c(r)^2-v(r)^2)}{r^2c(r)^2}~~~.\cr
 \end{align}

 We turn now to the general static spherically symmetric line element in S coordinates,
 \begin{equation}\label{Selt}
 ds^2=-F(r)^2 dt^2 + G(r)^2 dr^2 + r^2 d\Omega^2~~~,
 \end{equation}
 corresponding to $g_{tt}=-F(r)^2$ and $g_{rr}=G(r)^2$.
 Making again the convenient and symmetrical choice
 \begin{align}\label{nv2}
\ell^\nu=&\big(1/F(r),1/G(r),0,0\big) ~~~,\cr
n^\nu=&\big(1/F(r),-1/G(r),0,0\big)  ~~~,\cr
\end{align}
which also obey $\ell^\alpha  \ell_\alpha = n^\alpha  n_\alpha =0$ and $\ell^\alpha   n_\alpha =-2$, a Mathematica calculation  using the recipe of Eq. \eqref{expansion} gives\footnote{For a Schwarzschild black hole, $G(r)^2=1/F(r)^2=(1-2M/r)^{-1}$, so Eq. \eqref{Sexpan} becomes 
 $\theta_\ell \theta_n = -\frac{4}{r^2}(1-2M/r)$, giving the expected result that in this case the apparent horizon coincides with the event 
 horizon at $r=2M$.}   
                               
 \begin{align}\label{Sexpan}
 \theta_{\ell}=& \frac{2}{rG(r)}~~~,\cr
 \theta_{n}=& -\frac{2}{rG(r)}~~~,\cr
 \theta_\ell \theta_n =& -\frac{4}{r^2G(r)^2}~~~.\cr
 \end{align}

To map the (S) line element of Eq. \eqref{Selt} into GP form, one makes the change of variables
\begin{align}\label{map}
\hat{t} =&t-f(r)~,~~t=\hat{t}  +f(r), \cr
\hat{r}=&r~,~~ \hat{\Omega} = \Omega~~~,\cr
\end{align}
so that the line element becomes \big(with $ f^\prime(r)=df(r)/dr$\big)
\begin{align}\label{StoGP}
ds^2=&-F(r)^2\big(d\hat{t} +f^\prime(r)dr\big)^2 + G(r)^2 dr^2 + r^2 d\Omega^2~~~\cr
=&-F(r)^2 (d\hat{t} )^{\,2} -2F(r)^2 f^\prime(r) dr d\hat{t}  +[G(r)^2-F(r)^2 f^\prime(r)^2] dr^2 + r^2 d\Omega^2~~~.\cr
\end{align}
Choosing
\begin{equation}\label{fprimeformula}
f^\prime(r)=\frac{\big(G(r)^2-1\big)^{1/2}}{F(r)}
\end{equation}
transforms the S line element to GP form,
\begin{equation}\label{GPform}
ds^2=-[c(r)^2-v(r)^2](d\hat{t} )^{\,2}+2v(r)dr d\hat{t}   + dr^2 +r^2d\Omega^2~~~,
\end{equation}
with
\begin{equation}\label{cveqs}
 v(r)=-F(r)\big(G(r)^2-1\big)^{1/2}~,~~c(r)=F(r)G(r)~~~.
\end{equation}
From Eq. \eqref{GPexpan} we find now for $\theta_\ell$ and $\theta_n$
\begin{align}\label{GPexpan1}
 \theta_{\ell}=&\frac{2}{rG(r)}[G(r)+\big(G(r)^2-1\big)^{1/2}]=\kappa(r)\frac{2}{rG(r)} ~~~,\cr
 \theta_{n}=& -\frac{2}{rG(r)}[G(r)-\big(G(r)^2-1\big)^{1/2}]=\kappa(r)^{-1} \left(-\frac{2}{rG(r)}\right)~~~,\cr
 \theta_\ell \theta_n =& -\frac{4}{r^2G(r)^2}~~~,\cr
 \end{align}
 with $\kappa$(r) the scaling factor
 \begin{align}\label{kappaeq}
 \kappa(r)=&G(r)+\big(G(r)^2-1\big)^{1/2}~~~,\cr
 \kappa(r)^{-1}=&G(r)-\big(G(r)^2-1\big)^{1/2}~~~.\cr
 \end{align}

 Under the change of variables of Eq. \eqref{map}, the null vectors $\ell^{\nu}$ and  $n^\nu$ of Eq. \eqref{nv2} transform to
 \begin{align}\label{nv3}
\hat{\ell}_S^\nu=&\big(1/F(r)-f^\prime(r)/G(r),1/G(r),0,0\big) ~~~\cr
=&\left(\frac{1}{F(r)G(r)}\left(G(r)-\big(G(r)^2-1\big)^{1/2}\right),\frac{1}{G(r)},0,0\right)~~~,\cr
\hat{n}_S^\nu=&\big(1/F(r)+f^\prime(r)/G(r),-1/G(r),0,0\big)  ~~~\cr
=&\left(\frac{1}{F(r)G(r)}\left(G(r)+\big(G(r)^2-1\big)^{1/2}\right),-\frac{1}{G(r)},0,0\right)~~~.\cr
\end{align}
Multiplying $\hat{\ell}_S^\nu$ by $\kappa(r)$, and multiplying $\hat{n}_S^\nu$ by $\kappa(r)^{-1}$,
we have
\begin{align}\label{nv4}
\kappa(r)\hat{\ell}_S^\nu=&\big(1,c(r)-v(r),0,0\big)/c(r)=\ell_{GP}^\nu ~~~,\cr
\kappa(r)^{-1}\hat{n}_S^\nu=&\big(1,-c(r)-v(r),0,0\big)/c(r)=n_{GP}^\nu  ~~~,\cr
\end{align}
with $c(r)$ and $v(r)$  given by  Eq. \eqref{cveqs}.  Hence the rescaling of expansions $\theta_{\ell,\,n}$ in Eq. \eqref{GPexpan1} corresponds directly to the rescaling of null vectors  in Eq. \eqref{nv4}.

To conclude, although the expansions $\theta_\ell$ and $\theta_n$ transform as scalars under general coordinate transformations, as is explicit in the recipe of Eq. \eqref{expansion}, this can be deceptive in comparing results obtained by calculations in different coordinate systems.   Starting from a naturally convenient choice of the null vectors $\ell$ and $n$ in one coordinate system, one does not get the same expansions $\theta_{\ell,\,n}$ as calculated from a naturally convenient form of $\ell$ and $n$ in another coordinate system.  Instead, in general the calculational results for $\theta_{\ell,\,n}$ differ by a scale factor $\kappa(r)$ or its inverse, as a consequence of the covariance  of Eq. \eqref{expansion} under reciprocal rescalings of $\ell$ and $n$. Only the product $\theta_\ell \theta_n$, which can be used to locate the apparent horizon, is an invariant.

\end{document}